# Electronic Band Structures of Pristine and Chemically Modified Cellulose Allomorphs

Divya Srivastava,[1] Mikhail S. Kuklin,[1] Jouni Ahopelto,[2] Antti J. Karttunen[1,*]

[1] Department of Chemistry and Materials Science, Aalto University, P.O. Box 16100, FI-00076 Aalto, Finland (email: antti.j.karttunen@iki.fi)

[2] VTT Technical Research Centre of Finland Ltd, P.O. Box 1000, FI-02044 VTT, Espoo, Finland

We have investigated the structural properties, vibrational spectra, and electronic band structures of crystalline cellulose allomorphs and chemically modified cellulose with quantum chemical methods. The electronic band gaps of cellulose allomorphs $I_\alpha$, $I_\beta$, II, and $III_1$ lie in the range of 5.0 to 5.6 eV. We show that extra states can be created in the band gap of cellulose by chemical modification. Experimentally feasible amidation of cellulose $I_\beta$ with aniline or 4,4' diaminoazobenzene creates narrow bands in the cellulose band gap, reducing the difference between the occupied and empty states to 4.0 or 1.8 eV, respectively. The predicted states 4,4'diaminoazobenzene-modified cellulose $I_\beta$ fall in the visible spectrum, suggesting uses in optical applications.

## 1. Introduction

The increasing environmental, societal, and economical awareness have stimulated research towards the utilization and recovery of renewable bioenergy sources and environment-friendly functional materials. Cellulosic biomass is one of the material solutions when aiming towards more sustainable functional materials. Cellulose, the most abundant organic substance on earth, is found in plants, animals (tunicates), bacteria, algae, and fungi(Mohanty, Misra, & Hinrichsen, 2000)(Heinze, 2015)(Habibi, Lucia, & Rojas, 2010). However, the main source of cellulose is plant fiber: wood, bamboo, cotton, hemp, flax, jute, and other plant-based materials. The cell wall of all plants is mainly composed of cellulose, and it provides tensile strength, rigidity, and structure to the cell wall. In addition to being abundant, other properties such as nontoxicity, renewability, flexibility, biodegradability, light weight, transparency, and processability make cellulose and its derivatives potential materials for tackling industrial as well as environmental challenges. Future application areas for cellulose are envisaged in bioenergy, biotechnology, bio-composites, photonics and, potentially, in optoelectronics. (Mohanty et al., 2000)(Heinze, 2015)(Habibi et al., 2010)(Ingrao et al., 2015)(Berglund, 2005)(Nogi, Iwamoto, Nakagaito, & Yano, 2009)(Simão et al., 2015).

Cellulose is a linear homopolysaccharide polymer, in which β-D-anhdroglucopyranose units are linked by 1–4 glycosidic bonds (Figure 1 (a)). These D-glucopyranose rings within the cellulose chain prefer to be in a $^4C_1$ chair conformation, with all three (primary at C6 and two secondary at C2 and C3) hydroxyl groups in an equatorial positions and all hydrogen atoms in axial positions. The two ends of the cellulose chain are chemically different, the non-reducing end having an anomeric carbon bonded with glycosidic bonds while the reducing end has a D-glucopyranose unit in equilibrium with a cyclic hemiacetal. The three most probable conformations of hydroxymethyl groups about the C5-C6 bond are gauche-gauche (gg), trans-gauche (tg) and gauche-trans (gt). Newman projections of all three conformations are shown in Figure 1(b)(Perez & Mazeau, 2005).

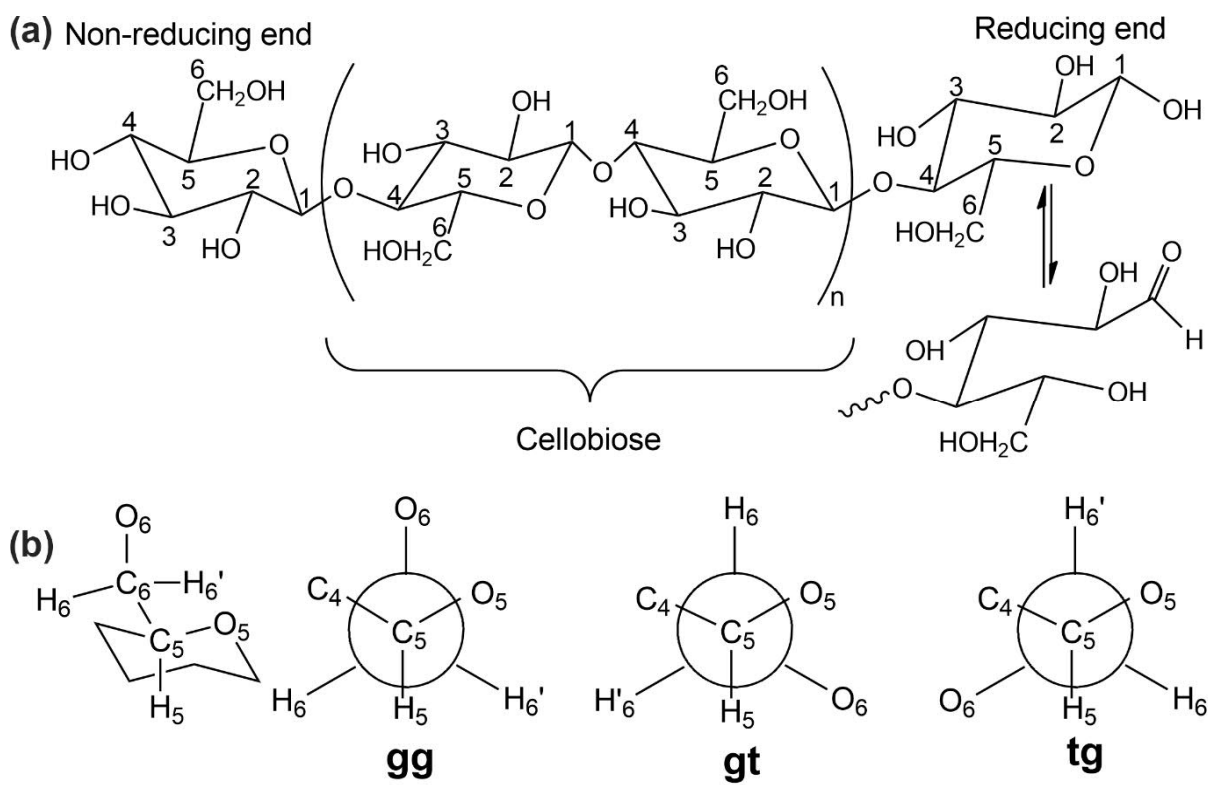

**Figure 1.** (a) Molecular structure of cellulose. (b) $^4C_1$ conformation of a D-glucopyranose ring and Newman projection of three most probable conformations of the hydroxymethyl groups in cellulose. Here, g is abbreviation of gauche (160°) and t is abbreviation of trans (180°) representing qualitatively the value of a dihedral angle. The first and second characters indicate the angle of $O_6$-$C_6$-$C_5$-$O_5$ moiety and $O_6$-$C_6$-$C_5$-$C_4$ moiety, respectively.

Important characteristics of cellulose are the crystallinity and the crystal symmetries. Six crystalline forms of cellulose have been identified. The native cellulose I is found in two crystalline forms, $I_\alpha$ and $I_\beta$, which both are usually co-present in cellulose. The ratio of $I_\alpha$ and $I_\beta$ within the sample depends on origin of the sample. Cellulose $I_\alpha$ is predominantly available in algae and bacterial cellulose, while cellulose $I_\beta$ is predominantly available in tunicin and higher plants. Nishiyama *et al.* used X-ray and neutron diffraction techniques to investigate the crystal structure of cellulose $I_\alpha$ and $I_\beta$(Nishiyama, Langan, & Chanzy, 2002; Nishiyama, Sugiyama, Chanzy, & Langan, 2003).

Cellulose $I_\alpha$ is triclinic (*P*1) and the unit cell contains one cellulose chain with all glucosyl linkage and hydroxymethyl groups being identical. $I_\beta$ is monoclinic (*P*2$_1$) with two conformationally distinct parallel chains (corner and center chains). In both cellulose I$_\alpha$ and I$_\beta$, all hydroxymethyl groups adopt trans-gauche conformation, and the structure contains sheets of cellulose chains stacked in a parallel-up fashion. The trans-gauche conformation allows the formation of intrachain hydrogen bonds and hydrogen bonds between the cellulose chains within a sheet. There are no O-H…O hydrogen bonds between the sheets and the sheets in I$_\alpha$ and I$_\beta$ are held together by only weak C-H…O interactions and van der Waals interactions. The main differences between I$_\alpha$ and I$_\beta$ structures are in the conformation of anhydroglucose residue, the glucosyl linkage, and their hydrogen bond network.

Other crystalline forms of cellulose, namely, II, III$_1$, III$_2$, IV$_1$, and IV$_2$ are synthesized through chemical processing of natural cellulose I. Cellulose II can be synthesized from cellulose I by two different processes: regeneration and mercerization(Habibi et al., 2010). In the regeneration process, cellulose I is dissolved in an appropriate solvent (heavy-metal-amine complexes), then reprecipitated in water. Mercerization process includes swelling of cellulose I in concentrated aqueous NaOH and cellulose II is produced after removing the swelling agent. Cellulose II has a monoclinic $P2_1$ crystal structure with two antiparallel chains having different conformations(Langan, Nishiyama, & Chanzy, 2001). Unlike cellulose I, in cellulose II the hydroxymethyl groups adopt gauche-trans conformation. Therefore, in cellulose II there is only one intrachain hydrogen bond between secondary OH at C3 and ring oxygen of another anhydroglucose residue (O3-H...O5), while in cellulose I, there are two intrachain hydrogen bonds O3-H...O5 and between OH at C2 and O at C6 (O2-H…O6). Cellulose II shows hydrogen bonds between the sheets.

Cellulose III$_1$ and III$_2$ can be obtained by exposing cellulose I and II in gas/liquid ammonia or various amines, respectively. Cellulose III$_1$ is monoclinic ($P2_1$) with one chain in the unit cell. In cellulose III$_1$, chains are parallel like in cellulose I, but hydroxymethyl groups are in gauche-trans conformation, and the inter-sheet hydrogen-bonding is similar to cellulose II(Wada, Chanzy, Nishiyama, & Langan, 2004; Wada, Nishiyama, Chanzy, Forsyth, & Langan, 2008). The crystal structure of cellulose III$_2$ is not well-established yet. Cellulose III$_1$/III$_2$ reverts to its parent cellulose I/II in high temperature and humid environment. Cellulose IV$_1$/IV$_2$ may be produced by heating cellulose III$_1$/III$_2$ up to 260C in glycerol(Habibi et al., 2010).

We use quantum chemical methods to study the crystal structure, electronic properties, and vibrational spectra of bulk cellulose I$_\alpha$, I$_\beta$, II, and III$_1$. The computational results are carefully compared to experimental data reported in the literature to find out the most suitable methodology to describe the structural and electronic properties of cellulose. Electronically, cellulose allomorphs are known to be insulators, but little is known from their actual electronic band structures. Our hypothesis is that crystalline cellulose can possess a dispersive band structure, opening up the possibility to tune the electronic and optical properties by controlled chemical modification. We investigate the chemical modification of cellulose by amidation, using slab models that reproduce the electronic properties of the bulk material.

## 2. Computational details

We carried out periodic density functional theory (DFT) calculations using both DFT-PBE generalized gradient approximation (GGA) exchange correlation functional(Perdew, Burke, & Ernzerhof, 1996) and DFT-PBE0 hybrid functional(Adamo & Barone, 1999). All calculations have been performed using the CRYSTAL17 program package(Dovesi et al., 2018). We used Gaussian-type triple-ζ-valence plus polarization level basis set (TZVP) derived from the molecular def2 basis set (basis set details are available in Supporting Information) (Weigend & Ahlrichs, 2005). The van der Waals-type dispersion interactions were described using Grimme's semiempirical DFT-D2 and DFT-D3 (zero-damping) dispersion corrections (Grimme, 2006)(Grimme, Antony, Ehrlich, & Krieg, 2010). In the case of bulk cellulose, the older DFT-D2 approach performs better in comparison to DFT-D3 when the experimental crystal structures are used as a benchmark. DFT-D3 underestimates the inter-sheet lattice constant by more than 4%, while DFT-D2 yields lattice constants with error less than 2% (see Table 1 in Supporting Information for detailed structural information of bulk cellulose allomorphs obtained with PBE-D3, PBE0-D3 and PBE0-D2 functionals). All calculations reported in the main paper were performed at the DFT-PBE-D2/TZVP level of theory.

The structures of the studied bulk cellulose allomorphs were fully optimized within their respective space groups. We modelled the chemical modification of cellulose $I_\beta$ by using two cellulose sheets parallel to (100) plane (2D bilayer slab model). While optimizing the adsorbates on the cellulose $I_\beta$ bilayer slab model, the bottom cellulose sheet was kept fixed, while the top layer adsorbate was optimized freely. The default CRYSTAL17 convergence criteria were used in the structural optimizations. Coulomb and exchange integral tolerances were set to tight values of 8, 8, 8, 8, and 16. The reciprocal space was sampled with Monkhorst-Pack $k$-meshes that are reported in SI (Pack & Monkhorst, 1977). Brillouin zone paths for the band structure diagrams were obtained from SeeK-path web service(Hinuma, Pizzi, Kumagai, Oba, & Tanaka, 2017).

The vibrational frequencies at the Γ-point were calculated within the harmonic approximation by evaluating the second derivatives of the potential energy with respect to atomic positions, as implemented in CRYSTAL code (Pascale et al., 2004)(Zicovich-Wilson et al., 2004). In the evaluation of Gibbs free energy, phonon $q$-sampling beyond Γ-point was not considered. Infrared (IR) and Raman intensities were calculated using the Coupled Perturbed Kohn-Sham method implemented in CRYSTAL(Maschio, Kirtman, Rérat, Orlando, & Dovesi, 2013c)(Maschio, Kirtman, Rérat, Orlando, & Dovesi, 2013a)(Maschio, Kirtman, Rérat, Orlando, & Dovesi, 2013b). The vibrational spectra were interpreted by visual inspection of the normal modes using the Jmol program package ("Jmol: An Open-Source Java Viewer for Chemical Structures in 3D. http://www.jmol.org/, Jmol Team," 2019).

## 3. Results

**3.1. Structures and energetics of bulk cellulose allomorphs.** The optimized lattice parameters of all studied cellulose allomorphs are listed in Table 1 and their optimized crystal structures are illustrated in Figure 2. For all cases, DFT-PBE with D2 dispersion correction describes the lattice parameters very well, showing in most cases differences smaller than 2% in comparison to experiments. The only exception is cellulose $III_1$, where the $c$ parameter is underestimated by 3.2%. The good performance of DFT-PBE-D2 for the cellulose allomorph structures is in line with the previous literature findings for allomorphs $I_\alpha$ and $I_\beta$ (Kubicki, Mohamed, & Watts, 2013).

The cellulose $I_\beta$ allomorph shows the lowest total energy and total Gibbs free energy (at 300 K) per cellulose chain. We define $\Delta E$ for each allomorph as:

$$\Delta E = E(\text{allomorph})/Z(\text{allomorph}) - E(\text{cellulose I}_\beta)/Z(\text{cellulose I}_\beta) \quad (1)$$

Where $E(\text{allomorph})$ is the total energy of a cellulose allomorph and $Z(\text{allomorph})$ is the number of cellulose chains in the unit cell. $\Delta G$ at 300 K is defined analogously to $\Delta E$, with $G(\text{allomorph})$ being defined as

$$G(\text{allomorph}) = E_e + E_0 + E_T + pV - TS, \quad (2)$$

where $E_e$ corresponds to total electronic energy, $E_0$ is the zero-point energy, $E_T$ is thermal contribution to vibrational energy, $p$ is pressure, $V$ is volume, $T$ is temperature, and $S$ is entropy. The $\Delta E$ and $\Delta G$ values listed in Table 1 show that the consideration of Gibbs free energy does not change the relative stability order of the cellulose allomorphs.

**Table 1.** Optimized structural parameters, indirect energy band gaps ($E_g$), and relative energies ($\Delta E$ and $\Delta G$) of crystalline cellulose allomorphs I$_\alpha$, I$_\beta$, II and III$_1$ at the DFT-PBE-D2/TZVP level of theory. The corresponding experimental values of the lattice parameters are reported in parentheses. The experimental references for cellulose allomprphs are given in the header row.

|  | I$_\alpha$ (Nishiyama et al., 2003) | I$_\beta$ (Nishiyama et al., 2002) | II (Langan et al., 2001) | III$_1$ (Wada et al., 2004) |
| --- | --- | --- | --- | --- |
| Crystal system | Triclinic | Monoclinic | Monoclinic | Monoclinic |
| Space group | $P1$ | $P2_1$ | $P2_1$ | $P2_1$ |
| $a$ (Å) | 6.03 (5.96) | 7.91 (7.78) | 8.01 (8.10) | 4.47 (4.45) |
| $b$ (Å) | 10.42 (10.40) | 10.42 (10.38) | 10.35 (10.31) | 10.40 (10.31) |
| $c$ (Å) | 6.62 (6.72) | 8.15 (8.20) | 9.09 (9.03) | 7.60 (7.85) |
| $\alpha$ (°) | 116.9 (118.1) | 90 | 90 | 90 |
| $\beta$ (°) | 114.9 (114.8) | 95.6 (96.5) | 111.9 (117.1) | 100.6 (105.1) |
| $\gamma$ (°) | 80.0 (80.4) | 90 | 90 | 90 |
| $E_g$ (eV) | 5.5 | 5.5 | 5.2 | 5.4 |
| $\Delta E$ (kJ/mol per $Z$) | 3.7 | 0 | 84.8 | 8.1 |
| $\Delta G$ (kJ/mol per $Z$) | 3.8 | 0 | 74.2 | 7.7 |

**3.2. Electronic structure of bulk cellulose allomorphs.** The electronic band structures and density of states of the studied cellulose allomorphs are shown in Figure 3. All the studied crystalline forms of cellulose are insulators with relatively flat bands, indicating localized charge carriers (see Table 1 for calculated band gaps). The band gap of 5.5 eV predicted for both I$_\alpha$ and I$_\beta$ allomorph is in good agreement with previous DFT band gaps of 5.7 eV for I$_\alpha$ and 5.4 eV for I$_\beta$ (Li, Lin, & Davenport, 2011). Even though the electronic bands are relatively flat, there are both valence and conduction bands showing some dispersion. The density of states at valence band maxima and conduction band minima show rather sharp features.

Because the DFT-PBE functional is known to underestimate the band gaps of insulators and semiconductors, we also calculated the band gap with hybrid PBE0 functional. The computed band gaps are 8.2, 8.1, 7.8, and 8.1 eV for cellulose I$_\alpha$, I$_\beta$, II, and III$_1$, respectively. The optical band gap of nanofibrillated cellulose is 4.5 eV based on absorption and cathodoluminescence

measurements(Simão et al., 2015). The band gap predicted by DFT-PBE is much closer to the experimental value compared to DFT-PBE0. To summarize, both the structural and electronic properties of cellulose obtained at the DFT-PBE-D2/TZVP level of theory are in line with the experimental data.

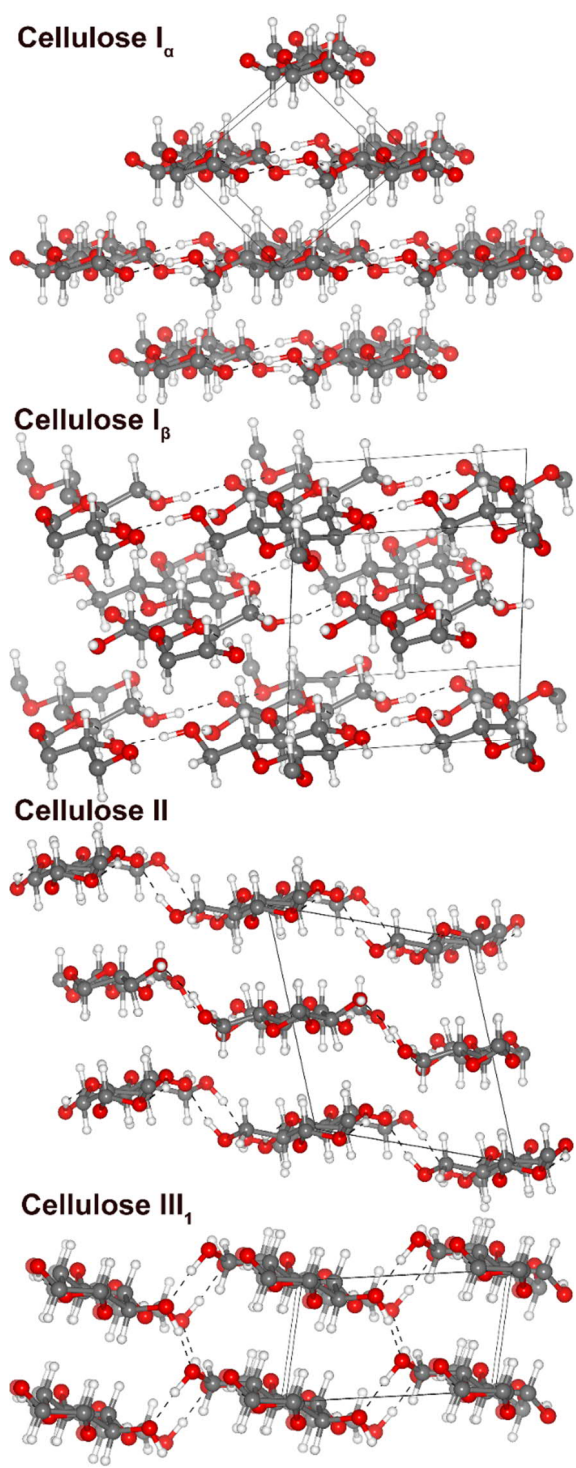

**Figure 1.** Optimized crystal structures of cellulose allomorphs $I_\alpha$, $I_\beta$, II, and $III_1$. Gray, red, and white spheres represent C, O, and H atoms, respectively. Hydrogen bonds are represented with dashed lines.

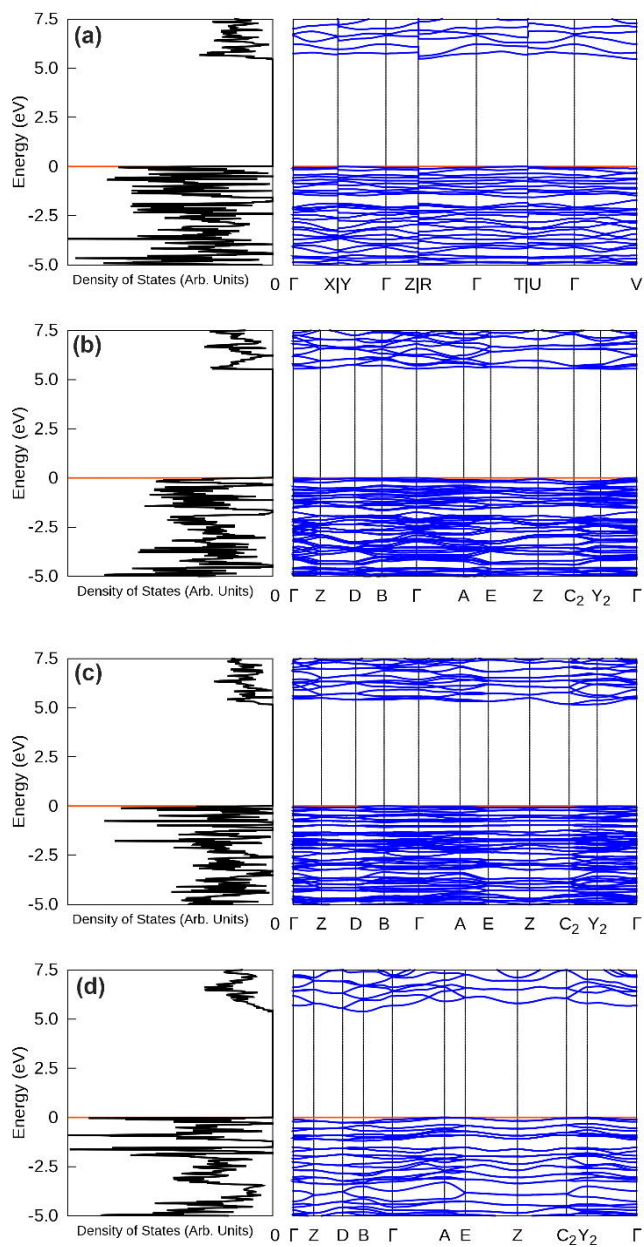

**Figure 2.** Electronic band structures and density of states for (a) cellulose $I_\alpha$, (b) cellulose $I_\beta$, (c) cellulose II, and (d) cellulose $III_1$. The top of the valence bands is at 0 eV.

### 3.3. Infrared and Raman spectra of bulk cellulose allomorphs.

Infrared (IR) and Raman specroscopy are powerful tools for investigating the structural characteristics of cellulose allomorphs. The predicted IR and Raman spectra for the studied cellulose allomorphs are shown in Figure 4. A detailed assignment of the vibrational bands is available in Table S3 of the Supporting Information, together with comparisons to previous literature (Lee et al. 2013, Marechal and Chanzy 2000). In the illustrated vibrational spectra, peaks in the lower wavenumber region (below 1500 cm$^{-1}$) are attributed to CH bending and peaks in region around 2800 to 3000 cm$^{-1}$ are assigned to CH stretching. Modes around 3000–3500 cm$^{-1}$ are attributed to OH stretching. OH vibrations in cellulose are highly coupled due to intra- and inter-chain hydrogen-bonding interactions.

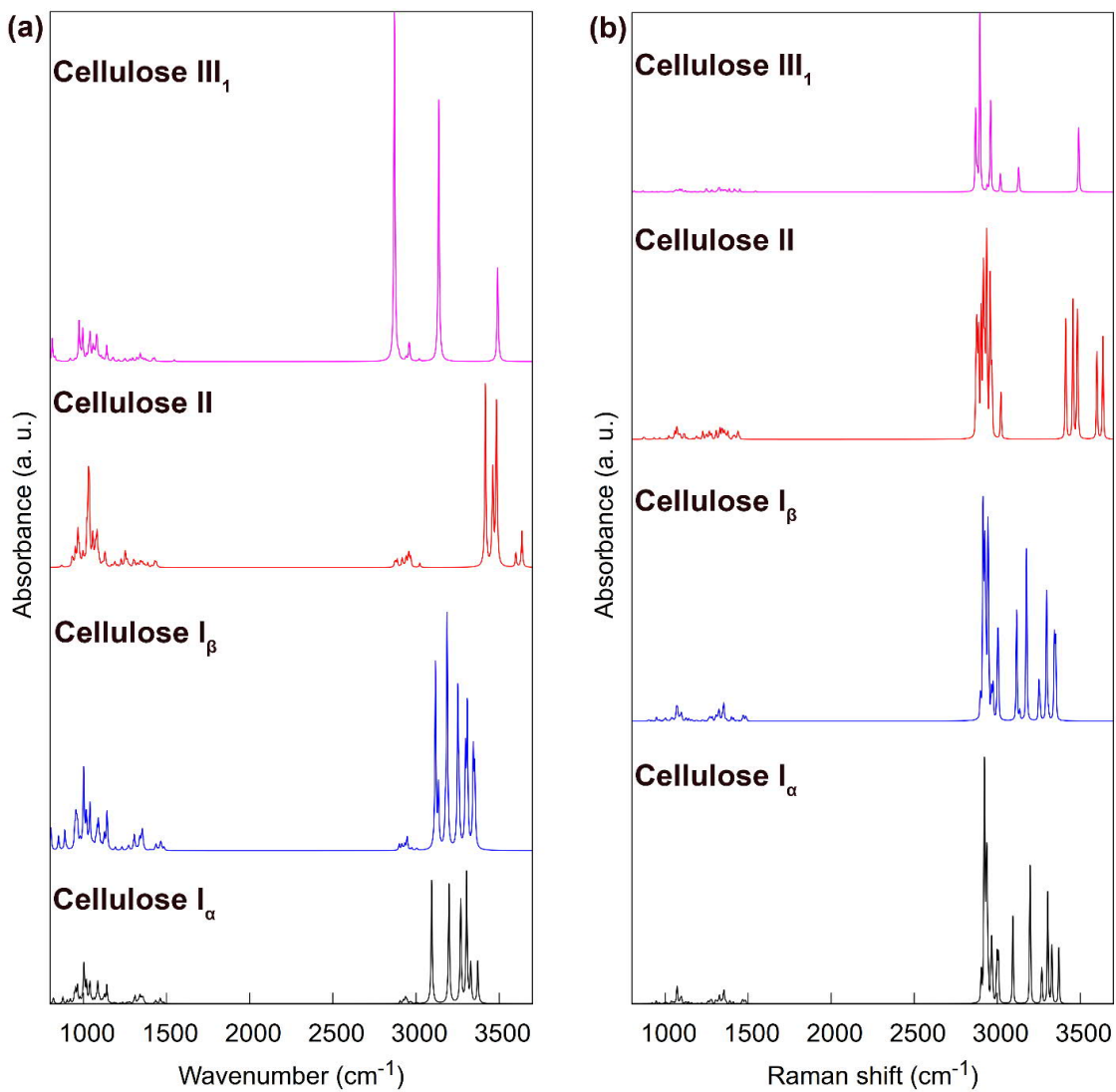

**Figure 4**. (a) Infrared and (b) Raman spectra of cellulose allomorphs $I_\alpha$, $I_\beta$, II, and $III_1$.

### 3.4 Geometry and electronic structure of chemically modified cellulose $I_\beta$

Even though the cellulose allomorphs are insulating materials with relatively large band gaps, chemical surface modification provides a way to tune their optoelectronic properties. Here, we investigate the electronic structure of pristine and chemically modified cellulose by using a bilayer (100) slab of the lowest energy allomorph cellulose $I_\beta$ (see Computational details). The electronic properties of the pristine bilayer slab model are very close to the bulk material (see below). This finding is in line with previous DFT-PBE calculations where it was observed that the surface energy and the electronic gap of cellulose $I_\beta$ (100) surface converged rapidly with the number of layers (Li et al., 2011).

We modify the cellulose $I_\beta$ with aniline ($C_6H_5NH_2$) and 4,4' diaminoazobenzene ($C_{12}H_{12}N_4$) molecules (two aniline rings joined *via* an azo group, N=N) to evaluate the effect of the chemical modification on the band structure.

The chemical surface modification approach adopted here is based on the experimentally known amidation approach of TEMPO-oxidized cellulose (Scheme 1 (Hakalahti et al., 2016)):

$$R-CH_2OH \xrightarrow{\text{TEMPO mediated oxidation}} R-\underset{\text{OH}}{\overset{\text{O}}{C}} \xrightarrow{R'-NH_2} R-\underset{\underset{H}{N}}{\overset{\text{O}}{C}}-R' + H_2O$$

**Scheme 1.** TEMPO-oxidation and amidation of cellulose.

The first step is to obtain TEMPO-oxidized cellulose,[30] followed by amidation with a primary amine (here aniline or 4,4'diaminoazobenzene). TEMPO-mediated oxidation can selectively oxidize primary alcohol groups of cellulose to carboxyl groups(Isogai et al., 2009) and here we only consider amidation at the C6 position of cellulose. A schematic representation of the chemical modification is shown in Figure 5. In the case of 4,4' diaminoazobenzene, only one of the amino groups forms a bond with the cellulose. Figure 6 displays the optimized structures resulting from the chemical surface modification. The amides bind to the cellulose surface with an angle of around 115° (C5-C6-N). For both molecules, the bond length between C6 and N is 1.36 Å and the C=O bond between C6 and O is 1.23 Å.

**Figure 5.** Schematic representation of bonding of aniline molecule and (b) 4,4' diaminoazobenzene molecule with C6 primary alcohol of cellulose.

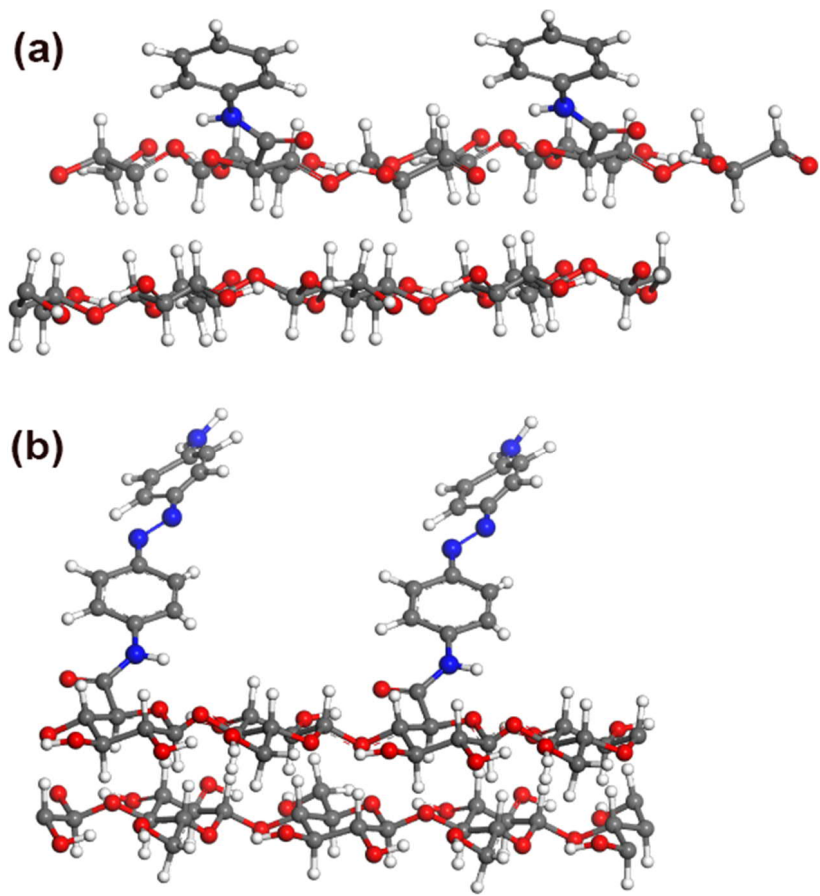

**Figure 6.** Chemical modification of TEMPO-oxided cellulose I$_\beta$ by amidation with (a) aniline and (b) 4,4' diaminoazobenzene.

Figure 7 shows the electronic band structure and density of states for pristine, aniline-modified and 4,4'diaminoazobenzene-modified cellulose I$_\beta$ (100) bilayer slab. The band gap in pristine cellulose I$_\beta$ (100) bilayer slab is 5.7 eV, in line with bulk cellulose I$_\beta$ (5.5 eV). The amidation introduces new electronic states into the band gap of cellulose. The gap between the occupied and empty states is reduced to 4.0 eV or 1.8 eV when the cellulose is modified with aniline or 4,4'diaminoazobenzene, respectively. The valence band maxima (VBM) and conduction band minima (CBM) of the pristine system occur at -5.81 and -0.14 eV, respectively. In the case of aniline, VBM and CBM occur at -5.46 and -1.5 eV, while for 4,4'diaminoazobenzene VBM is -4.68 and CBM is -2.82 eV.

The atom-projected density of states in Figure 8 illustrate clearly how the aniline and 4,4'diaminoazobenzene introduce new states into the band gap of cellulose. In the aniline-modified system, the lowest-energy conduction band arises from C and N atoms of the aniline, while the band around 4.5 eV is due to the aniline phenyl ring only. In 4,4'diaminoazobenzene-modified system, the states around 1.8 eV and 3.3 eV are due to the azo group (N=N) and $C_6H_4NH_2$ ring, respectively. The band structures of the modified cellulose indicate that the band gap of cellulose can be tuned through well-controlled chemical modification.

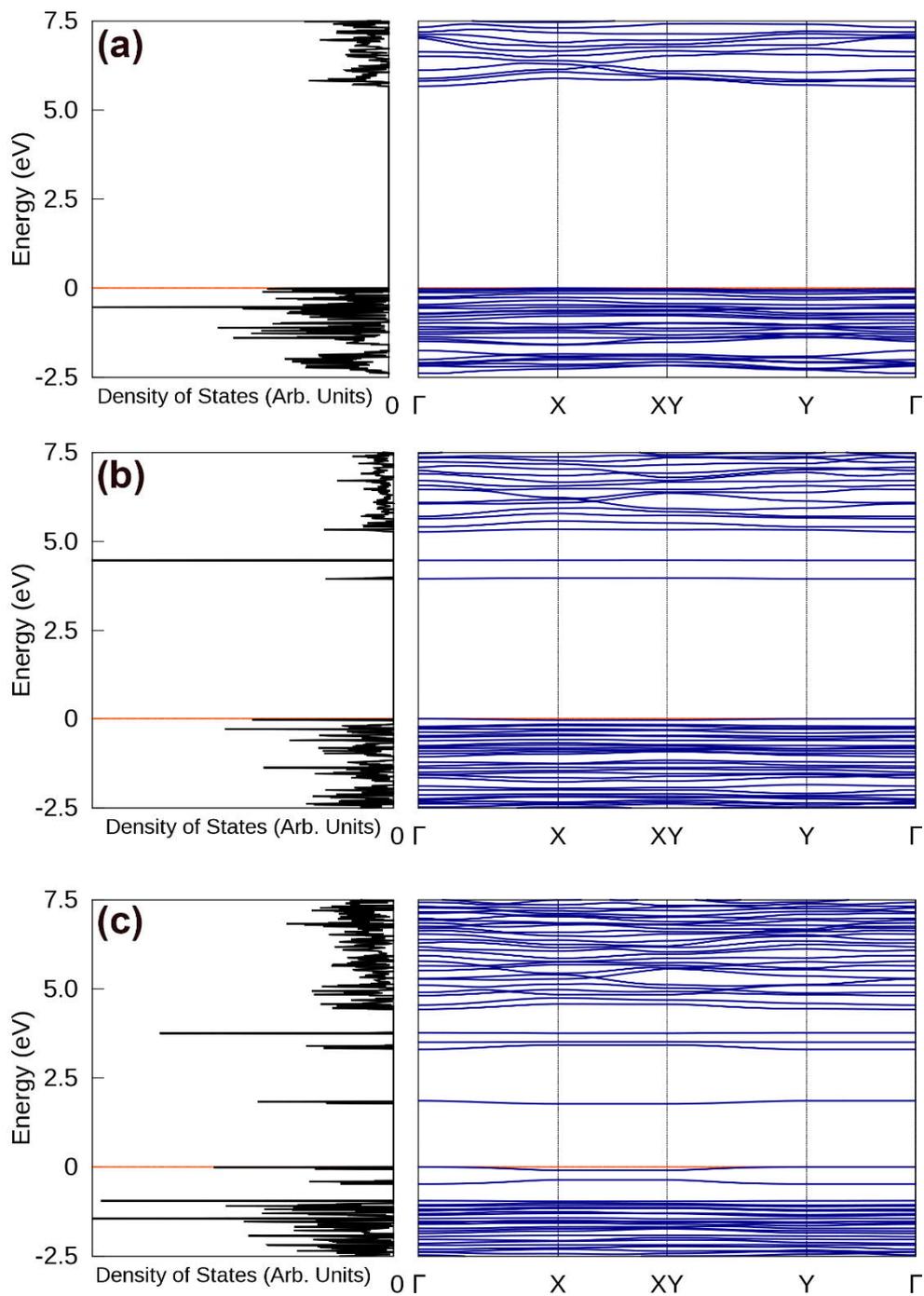

**Figure 7.** Electronic band structure of (a) pristine, (b) aniline-modified and (c) 4,4'-diaminozaobenzene-modified cellulose I$_\beta$ (100) bilayer slab. The top of the valence bands is at 0 eV.

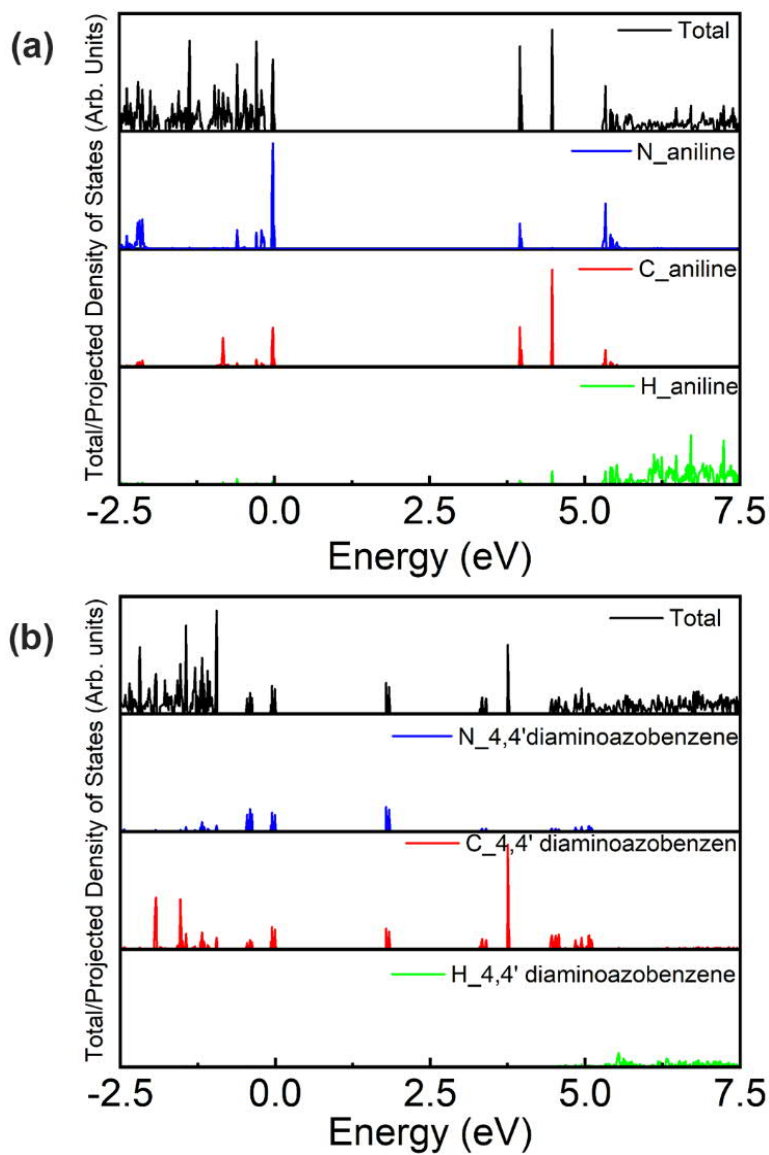

**Figure 8.** Total density of states (DOS) and atom-projected DOS for (a) aniline-modified and (b) 4,4'diaminoazobenzene-modified cellulose I$_\beta$ (100) bilayer slab. The top of the valence bands is at 0 eV.

## 4. Conclusion

We have investigated the structural, electronic, and vibrational properties of bulk cellulose allomorphs I$_\alpha$, I$_\beta$, II and III$_1$ at the DFT-PBE-D2/TZVP level of theory, which provides a good description of the properties when compared to experimental data. Cellulose I$_\beta$ is the lowest-energy allomorph both in terms of electronic total energy and Gibbs free energy. The predicted band gaps of I$_\alpha$, I$_\beta$, II and III$_1$ allomorphs are 5.5, 5.5, 5.2 and 5.4 eV, respectively. We showed that the band structure of cellulose can be modified by experimentally feasible chemical modifications. Chemical modifications of cellulose I$_\beta$ with aniline and 4,4'diaminoazobenzene create states at energies of 4.0 or 1.8 eV above the valence band, respectively. Our findings demonstrate that the optical absorption of cellulose can be tuned from ultraviolet to visible range with controlled amidation reactions, making nanocellulose an interesting material platform for photonics applications.

## Acknowledgments

We thank Maisa Vuorte and Dr. Maria Sammalkorpi for helpful discussions on the atomic-level structural details of cellulose. This work was a part of the Academy of Finland's Flagship Programme under Projects No. 318890 and 318891 (Competence Center for Materials Bioeconomy, FinnCERES). Computational resources were provided by CSC, the Finnish IT Center for Science.

Supporting Information

# Electronic Band Structures of Pristine and Chemically Modified Cellulose Allomorphs

Divya Srivastava,[1] Mikhail S. Kuklin,[1] Jouni Ahopelto,[2] Antti J. Karttunen[1,*]

[1] Department of Chemistry and Materials Science, Aalto University, P.O. Box 16100, FI-00076 Aalto, Finland (email: antti.j.karttunen@iki.fi)

[2] VTT Technical Research Centre of Finland Ltd, P.O. Box 1000, FI-02044 VTT, Espoo, Finland

# Structural parameters of cellulose allomorphs with DFT-PBE-D3, DFT-PBE0-D3, and DFT-PBE0-D2

Table S1 shows the lattice parameters of cellulose allomorphs $I_\alpha$, $I_\beta$, II, and III$_1$ optimized with DFT-PBE0-D3, DFT-PBE0-D2, and DFT-PBE-D3 methods (TZVP basis set). DFT-D3 (zero-damping) leads in larger differences with experiments in comparison to DFT-D2.

**Table S1.** Lattice parameters of cellulose $I_\alpha$, $I_\beta$, II, and III$_1$ allomorphs obtained with DFT-PBE0-D3, DFT-PBE0-D2, and DFT-PBE-D3 methods. The numbers in parentheses show the difference to experimental values.

| Parameter | $I_\alpha$ | $I_\beta$ | II | III$_1$ |
|---|---|---|---|---|
| DFT-PBE0-D3/TZVP | | | | |
| $a$ (Å) | 5.79 (2.85%) | 7.40 (4.88%) | 7.83 (3.33%) | 4.33 (2.70%) |
| $b$ (Å) | 10.31 (0.87%) | 10.32 (0.58%) | 10.36 (0.48%) | 10.26 (0.48%) |
| $c$ (Å) | 6.57 (2.23%) | 8.10 (1.22%) | 8.45 (6.42%) | 7.57 (3.57%) |
| $\alpha$ (°) | 118.38 | 90.00 | 90.00 | 90.00 |
| $\beta$ (°) | 115.26 | 96.71 | 114.43 | 101.78 |
| $\gamma$ (°) | 81.56 | 90.00 | 90.00 | 90.00 |
| Band gap (eV) | 8.1 | 8.3 | 8.2 | 8.1 |
| DFT-PBE0-D2/TZVP | | | | |
| $a$ (Å) | 6.00 (0.67%) | 7.88 (1.28%) | 7.95 (1.88%) | 4.47 (0.45%) |
| $b$ (Å) | 10.34 (0.58%) | 10.34 (0.39%) | 10.28 (0.29%) | 10.30 (0.1%) |
| $c$ (Å) | 6.66 (0.89%) | 8.15 (0.61%) | 9.10 (0.78%) | 7.75 (1.27%) |
| $\alpha$ (°) | 117.22 | 90.00 | 90.00 | 90.00 |
| $\beta$ (°) | 114.99 | 96.27 | 116.34 | 103.85 |
| $\gamma$ (°) | 79.70 | 90.00 | 90.00 | 90.00 |
| Band gap (eV) | 8.1 | 8.2 | 7.8 | 8.1 |
| DFT-PBE-D3/TZVP | | | | |
| $a$ (Å) | - | 7.52 (3.34%) | - | - |
| $b$ (Å) | - | 10.39 (0.1%) | - | - |
| $c$ (Å) | - | 8.11 (0.49%) | - | - |
| $\alpha$ (°) | - | 90.00 | - | - |
| $\beta$ (°) | - | 96.18 | - | - |
| $\gamma$ (°) | - | 90.00 | - | - |
| Band gap (eV) | | 5.5 | | |

**Table S2.** Reciprocal space $k$-meshes for all studied structures

| System | $k$-mesh |
|---|---|
| Cellulose $I_\alpha$ | 3×4×4 |
| Cellulose $I_\beta$ | 3×3×3 |
| Cellulose II | 3×4×4 |
| Cellulose III$_1$ | 8×5×3 |
| Cellulose I$\beta$ (100) bilayer slab | 3×3 |
| Cellulose I$\beta$ (100) bilayer slab with aniline | 3×3 |
| Cellulose I$\beta$ (100) bilayer slab with 4,4'diaminoazobenzene | 3×3 |

## Basis set details

The TZVP basis sets derived from molecular Karlsruhe def2 basis sets were taken from previous studies:

**O**, **C**, and **H**: (Karttunen, Tynell, & Karppinen, 2015)

**N:** (Ivlev, Müller, Karttunen, Hoelzel, & Kraus, 2018)

Ivlev, S. I., Müller, T. G., Karttunen, A. J., Hoelzel, M., & Kraus, F. (2018). A Neutron Diffraction and Quantum-Chemical Study of [Mn(ND3)6](N3)2. *Zeitschrift Fur Anorganische Und Allgemeine Chemie*, *644*(21), 1349–1353. https://doi.org/10.1002/zaac.201800217

Karttunen, A. J., Tynell, T., & Karppinen, M. (2015). Atomic-level structural and electronic properties of hybrid inorganic-organic ZnO:Hydroquinone superlattices fabricated by ALD/MLD. *Journal of Physical Chemistry C*, *119*(23), 13105–13114. https://doi.org/10.1021/acs.jpcc.5b03433

## Vibrational band assignments

**Table S3.** Summary of band assignments for the IR and Raman spectra of cellulose allomorphs.

| Frequency (cm$^{-1}$) | Assignment |
|---|---|
| **Cellulose I$\alpha$** | |
| < 1500 | C-H bending |
| 1003 | IR: The most intense peak in this region due to C-H bending |
| 2904-3007 | C-H stretching region, agrees reasonably well with experimetal region from 2855-2968.(Lee, Mohamed, Watts, Kubicki, & Kim, 2013) |
| 2923 | Major peak in Raman spectra, in experimental spectra at 2898 cm$^{-1}$ (Lee et al., 2013) |
| 2938 | Major peak in IR spectra, in experimental spectra at 2904 cm$^{-1}$ (Lee et al., 2013).[b] |
| 2938-2945 | C-H2 stretching coupled with C-H stretching |
| 2998-3007 | C-H2 stretching |
| 3198 to 3371 | OH stretch vibration frequencies region, is very close to the experimetally observed region 3200-3600 cm$^{-1}$ in IR and 3200-3400 cm$^{-1}$ in SFG (Lee et al., 2013). Both calculated IR and Raman spectra show six bands in this range. |
| 3094 | Mode due to mainly O2-H…O6 stretcing with small contribution from O3-H…O5. Intrachain O2-H…O6 interaction is the strongest hydrogen bond. |
| 3198-3268 | Inter-chain interaction O6-H…O3 and intrachain O3-H…O5 and O2-H...O6. |
| 3304 | Strongest OH peak due to interchain O6-H…O3 and intrachain O3-H…O5 interactions. |
| 3329 and 3371 | Assigned to O3-H…O5 and O6-H..O3 interactions |
| **Cellulose I$\beta$** | |
| 700-1050 | O-H bending of primary and secondary alcohals coupled with CH bending and CO stretching. |
| 1001 | Peak in IR corresponds mainly C6..O5 stretching along with C2..O2and C3…O3 stretching. |
| 1300-1500 | C-H and C-H2 bending |
| 1069 and 1323 | The calculated Raman Peaks in this region are close to experimentally observed Raman Peaks at 1098 and 1320 cm$^{-1}$ (Maréchal & Chanzy, 2000) |
| 2898-3000 | C-H and C-H2 stretching |
| 2947 | Strongest peak in this region due to C-H and C-H2 stretching, very close to experimentally observed peak at 2944 cm$^{-1}$ in SFG spectra of cellulose I$\beta$ (Lee et al., 2013). |
| 3100 to 3555 | O-H stretching region, closer to experimentally observed value 3200-3600 cm$^{-1}$ in IR(Lee et al., 2013)(Maréchal & Chanzy, 2000). The calculated IR and Raman spectra have five distinct peaks. |
| 3100-3258 | Peaks in this region have contribution mainly from intrachain O2-H...O6 interaction. As in cellulose I$\alpha$, O2-H…O6 interactions forms strongest hydrogen bond in cellulose I$\beta$. |
| 3297-3354 | Coupled contribution from intrachain O3-H..O5 and interchain O6-H…O3 interactions. |
| 3344 | Contribution from O3-H..O5 interactions, closer to experimetally observed peak at 3340 cm$^{-1}$ in I$\beta$ cellulose in *Valonia* ((Maréchal & Chanzy, 2000) see Table 1).[b] |
| **Cellulose II** | |
| 800-1500 | Raman spectra shows minor peaks compared to IR spectra in this regime. It corresponds to C-H and C-H2 bending along with O-H bending. The IR peaks at lower wavenumber correspond CH and secondary OH bending. |
| 1029 | strongest peak in this region, corresponds to bending modes of O-H and C-H along with stretching modes of C-O bond. |
| 1438-1447 | C-H2 bending |
| 2871 – 3022 | C-H and C-H2 stretching region, Raman spectra shows major peaks in this frequency range. Peaks around lower wavenumber region are CH stretching modes and around larger wavenumber correspond CH2 stretching |
| 2958 | Strongest peak in IR, mainly CH2 stretching with small contribution from CH stretching. |
| 2937 | Strongest peak in Raman spectra. |
| 3412-3638 | O-H stretching region having five distinct peaks |
| 3417 | Strongest peak in IR, assigned to O6-H…O2 inter-chain stretching. |
| 3456 | Strongest peak in Raman due to mainly inter-chain O6-H…O2 with small contribution from coupled intrachain O3-H…O6 and O3-H…O5 iteractions. |
| 3462-3485 | Mainly from inter-chain O6-H…O2 with intra-chain O3-H..O5, O3H…O6, and with little contribution from inter-sheet O2-H…O´6 and O6-H…O´2 hydrogen bonding networks. |
| 3600 and 3636 | Coupled interchain O6-H…O2 and O2-H…O6 interactions. |
| **Cellulose III₁** | |
| 800-15000 | Region dominated by C-H bending along with O-H bending and C-O and C-C stretching. Peaks around larger wavenumber range are due to C-H bending |
| 2869 to 3501 | Five distinct peaks in IR and six distinct peaks in Raman spectras. Vibration modes in this frequency range are due to C-H and O-H stretchings. |
| 2870-2940 | Modes in this range are due to C-H streching |
| 2870 and 2895 | Most intense peak in IR and Raman spectra, respectively due to C-H stretching. |
| 2960 | C-H and C-H2 stretchings |
| 3019 | C-H2 stretching. |
| 3129-3501 | O-H stretching region |
| 3129 and 3136 | Peaks in Raman and IR spectra, respectively corresponds coupled stretching of inter-chain and inter-sheet O6-H…O2 and O2-H…O6 hydrogen bonds. |
| 3491 | Peaks around this frequency range correspond to intra-chain O3-H…O5 hydrogen bonds stretching.[c] |

[a] Previous DFT-PBE-D2 calculations show peaks for C-H stretch modes around 2900 cm$^{-1}$(Lee et al., 2013).
[b] Our IR spectra in O-H stretching region is closer to experimental IR spectra obseved for cellulose I$_\beta$ in Valonia(Maréchal & Chanzy, 2000). The discrepancy between DFT-D2 and experimental peak positions in O-H stretching region may be due to small difference in opimized and expermental geometry of cellulose I$_\beta$.
[c] In cellulose III, interchain and inter-sheet O6-H..O2 and O2-H..O6 interactions form strong hydrogen bonding and intrachain O3-H..O5 interaction forms weak hydrogen bonding network.

## Optimized geometries of the studied structures

Lattice parameters and atomic positions of the studied structures are given below in CRYSTAL input format (DFT-PBE-D2/TZVP level of theory).

### Cellulose I$_\alpha$

```
P 1
10.41553208 6.62354735 6.02782654 79.968353 116.903941 114.925502
42
1    -1.459436037967E-01  -2.465697251151E-01   2.276479838935E-01
1     3.051897201667E-01  -3.494985035259E-01   3.580709824519E-01
1    -1.958507130822E-01   3.424385528262E-01  -3.503176188718E-01
1    -4.496215411835E-02   4.973551010398E-01   4.408810566652E-01
1     4.915936819043E-01  -4.824669591791E-01  -4.173870748405E-01
1     3.749849322975E-01   2.495030601135E-01  -2.021511718273E-01
8     2.390244265818E-01   1.548655464621E-01  -5.809099061446E-02
6     2.550851967611E-01  -5.395138000190E-02   3.279750796472E-02
1     1.954209615955E-01  -1.705246914932E-01  -1.276617131002E-01
8     4.141367754998E-01  -8.378773275407E-03   1.545488561162E-01
6     1.861645903034E-01  -1.549345417079E-01   2.236096886200E-01
1     2.553425989072E-01  -3.596079364480E-01   3.836596739761E-01
6     1.610729227104E-02  -1.804909736283E-01   1.092641841997E-01
1    -5.559003214083E-02  -3.104108622883E-01  -3.940426799788E-02
6     2.662180311293E-03   3.994410804537E-02  -2.971328365798E-03
1     6.182776469471E-02   1.659255857157E-01   1.474776616247E-01
6     7.804484967232E-02   1.313621607698E-01  -1.867482422954E-01
1     1.537246136435E-02   1.622935023962E-02  -3.488100951535E-01
6     8.368096512066E-02   3.648996362859E-01  -2.743596069998E-01
1     1.180567894220E-01   4.634516742675E-01  -1.116524497934E-01
1     1.731230262637E-01   4.419711628788E-01  -3.473385899116E-01
8     1.935585520657E-01  -3.670568221507E-01   2.925174082519E-01
8    -3.943497434710E-02  -2.466697389737E-01   3.009929731509E-01
```

```
8     -6.410286400257E-02    3.549580055923E-01   -4.676071578395E-01
8     -3.085863559170E-01   -1.550213838893E-01    7.474598289669E-02
6     -2.288730395141E-01    5.432520847785E-02   -1.737693703547E-02
1     -1.422528969766E-01    1.722932092061E-01    1.417026565752E-01
8     -1.582401571655E-01    8.997181094753E-03   -1.403225462683E-01
6     -3.442095888662E-01    1.539023307587E-01   -2.085607107781E-01
1     -4.243439981923E-01    3.494187920131E-02   -3.676345348651E-01
6     -4.374090200092E-01    1.827760279397E-01   -9.003100050311E-02
1     -3.568187540620E-01    3.077506041006E-01    6.355722303542E-02
6      4.882423938068E-01   -3.883680026689E-02    1.928616802701E-02
1      4.014239388955E-01   -1.616548476212E-01   -1.345688471045E-01
6     -3.895225600592E-01   -1.309898556128E-01    2.035285427639E-01
1     -3.063444614139E-01   -1.516642121562E-02    3.646534213975E-01
6     -4.615962721279E-01   -3.636590744601E-01    2.940965809139E-01
1      4.414442784424E-01   -4.724894693125E-01    1.334983685907E-01
1     -3.712396268339E-01   -4.335951477422E-01    3.828007951237E-01
8     -2.593889103321E-01    3.644704317844E-01   -2.805575422863E-01
8      4.447610279086E-01    2.556608543662E-01   -2.746966087989E-01
8      4.859615787012E-01   -3.534949060612E-01    4.723113190119E-01
```

**Cellulose I$_\beta$**

```
P 21
8.15005949 10.41600700 7.90704681 95.597247
42
6     -3.888072557482E-02    4.267977175950E-02    1.369986506461E-02
1     -1.334208129758E-02    5.567250091672E-02    1.530354730026E-01
6     -1.851384017598E-01   -4.921481236320E-02   -2.574280162219E-02
1     -2.098067590876E-01   -5.496096297317E-02   -1.654681654332E-01
6     -1.388518250488E-01   -1.816739950769E-01    4.577706966632E-02
1     -1.254756625266E-01   -1.780669181689E-01    1.862215355518E-01
6      2.563230111510E-02   -2.249510925552E-01   -1.525826276908E-02
1      8.896459805512E-03   -2.389895548816E-01   -1.545628876780E-01
6      1.629849642403E-01   -1.253051196708E-01    2.667609223426E-02
1      1.855088674606E-01   -1.130130451322E-01    1.660889519008E-01
6      3.251182954560E-01   -1.586415723313E-01   -4.572943267013E-02
1      3.002334908617E-01   -1.902402747733E-01   -1.790507409293E-01
1      4.010479199185E-01   -7.102613947583E-02   -4.363683664619E-02
8     -3.230956242347E-01   -1.544937629844E-03    5.187176500091E-02
1     -3.490585537417E-01    8.566733479149E-02    4.737723549865E-03
8     -2.671710050516E-01   -2.710834421091E-01   -1.047120281963E-02
1     -2.290473226293E-01   -3.578340829158E-01    2.637431453847E-02
8      7.983359312358E-02   -3.419241767337E-01    6.799150287891E-02
8      1.089293988778E-01   -5.517671850432E-03   -5.189693289132E-02
8      4.129169564380E-01   -2.554270899881E-01    5.510976655993E-02
1     -4.690893978927E-01   -2.493833001037E-01    3.270765977757E-02
6      4.579868305665E-01    3.057533654223E-01   -4.842862186513E-01
1      4.703675513751E-01    3.192716074660E-01   -3.446581498797E-01
6      3.158471319223E-01    2.134064898725E-01    4.617360805459E-01
1      3.044906776685E-01    2.079591059187E-01    3.211409465585E-01
6      3.573119092397E-01    8.070209177640E-02   -4.651104991634E-01
1      3.637659532193E-01    8.642210632016E-02   -3.248417092857E-01
6     -4.737370497211E-01    3.768493835158E-02    4.855239289765E-01
1     -4.798575163838E-01    2.391188543343E-02    3.464186782477E-01
6     -3.379999297286E-01    1.362449012687E-01   -4.614764094341E-01
1     -3.197119576410E-01    1.442278668341E-01   -3.216950097784E-01
6     -1.749545295225E-01    1.054456632926E-01    4.669669161088E-01
1     -2.023077776705E-01    8.137097729647E-02    3.308061126276E-01
1     -9.720573997636E-02    1.919429765297E-01    4.790173935241E-01
8      1.693761149756E-01    2.583624609361E-01   -4.757905892997E-01
1      1.503548295449E-01    3.495917206413E-01    4.872133557765E-01
8      2.299661964280E-01   -7.159179136254E-03    4.722288640632E-01
1      2.677505804488E-01   -9.494970236000E-02   -4.953860830739E-01
8     -4.245708178448E-01   -7.954227281233E-02   -4.287842131910E-01
8     -3.874790887249E-01    2.586734552534E-01    4.640812800224E-01
8     -9.034227655338E-02    2.913228088450E-03   -4.416537319125E-01
1      2.841820996228E-02    1.224489487222E-02   -4.567224825707E-01
```

**Cellulose II**

```
P 21
8.00593100 10.35360818 9.09438874 111.936264
42
6      5.424072804987E-02    3.567053180790E-01    1.664500211092E-02
1      3.811626416431E-02    3.627400701062E-01   -1.098634413178E-01
6      2.010601269373E-01    2.595929064089E-01    1.056586574777E-01
1      2.090350521442E-01    2.523338436572E-01    2.293514291841E-01
6      1.530684276081E-01    1.268323808694E-01    2.718545142979E-02
1      1.551674927371E-01    1.336199410333E-01   -9.399828966959E-02
8      3.713617762402E-01    2.987699099329E-01    1.024945839182E-01
6     -4.044412018212E-02    8.945699535175E-02    9.046931389641E-03
1     -3.877323581663E-02    6.720212488728E-02    1.284888875637E-01
8      2.815639865948E-01    3.887136735793E-02    1.232486894521E-01
6     -1.776215014390E-01    1.982025354474E-01   -6.431367451283E-02
1     -1.910295746346E-01    2.155920580634E-01   -1.883568949561E-01
8     -1.006584284385E-01   -2.300549885419E-02   -9.064255781457E-02
8     -1.117534950969E-01    3.140165982339E-01    2.739970641867E-02
6     -3.616002922642E-01    1.674358561448E-01   -5.923692258840E-02
1     -4.098306788513E-01    7.785282255137E-02   -1.280837785923E-01
1     -3.445559222532E-01    1.481132436264E-01    6.513633901039E-02
8     -4.839962226736E-01    2.696277284123E-01   -1.248824324555E-01
6      6.130726784279E-02   -1.253251018416E-01   -4.733220683937E-01
1      6.541805208703E-02   -1.345015510445E-01   -3.502152673281E-01
6      2.024863703351E-01   -2.855385112932E-02   -4.801261967886E-01
1      2.083358301332E-01   -3.337352925438E-02    4.004378489496E-01
6      1.609958729480E-01    1.122952867253E-01   -4.505006027397E-01
1      1.972513259689E-01    1.261108094946E-01   -3.217606110932E-01
8      3.758135628624E-01   -6.477707219358E-02   -3.634492317630E-01
6     -4.390232060226E-02    1.419020750416E-01    4.659420917139E-01
1     -7.333037738931E-02    1.554536395302E-01    3.380745458641E-01
8      2.686681431775E-01    1.911027084502E-01    4.927792810097E-01
6     -1.594648518340E-01    3.121244427912E-02    4.878800861476E-01
1     -1.291178084489E-01    1.734847354062E-02   -3.837672246676E-01
8     -9.924531350873E-02    2.556346430434E-01   -4.728408886673E-01
8     -1.107175073893E-01   -8.309155394982E-02    4.238450298945E-01
6     -3.621472363496E-01    4.608691686375E-02    4.035345146521E-01
1     -4.088177818903E-01    1.334379992025E-01    4.460793350931E-01
1     -3.973733159545E-01    5.543627016959E-02    2.749455612580E-01
8     -4.481378986763E-01   -6.615844974040E-02    4.305778098519E-01
1     -3.089661222470E-01   -2.301225848162E-01    4.428470513904E-01
1      3.798069260516E-01   -1.591292739199E-01   -3.673665088452E-01
1      4.873872305011E-01    4.514766061223E-01    4.812167854373E-01
1      4.494156400284E-01    2.826353253701E-01   -5.427418307695E-02
1      3.963018475731E-01    3.853467650818E-01    1.478382486692E-01
1      2.662614380871E-01   -4.550719633459E-02    7.027700031809E-02
```

**Cellulose III$_1$**

```
P 21
4.47339236 7.60434907 10.40458253 100.609050
21
6      1.454441779658E-02    5.438926217999E-02    3.832151494485E-01
6      1.702752799206E-01    1.941950900608E-01    2.871253265551E-01
6      3.111011900531E-02    1.642265476252E-01    1.517826698344E-01
6     -2.208509526944E-03   -3.421849952294E-02    1.135784212767E-01
6     -1.604888691643E-01   -1.578359036612E-01    2.194835942977E-01
6     -1.845544827109E-01   -3.554446653952E-01    1.864897883462E-01
8      1.600727642481E-01    3.673096038634E-01    3.353079993574E-01
8      2.300619867108E-01    2.790830446231E-01    6.899577178642E-02
8     -1.770344150106E-01   -7.365209671470E-02   -2.180845576046E-03
8      2.089222245577E-02   -1.217021729569E-01    3.338018253407E-01
8     -2.903651394398E-01   -4.626080647124E-01    2.958307919858E-01
1     -2.256977058464E-01    6.799756960734E-02    3.990108164291E-01
1      4.093131171535E-01    1.767199798607E-01    2.798026086464E-01
1     -1.998223754626E-01    1.981157592215E-01    1.516068469142E-01
1      2.278443947192E-01   -6.453058765467E-02    9.888208519484E-02
1     -3.908840199837E-01   -1.293539745635E-01    2.386400676321E-01
1     -3.399547610185E-01   -3.890938203399E-01    1.045100379918E-01
```

```
   1      3.984352216666E-02   -3.819965237396E-01    1.571409280799E-01
   1      3.657017341340E-01    4.439660906671E-01    3.184534190583E-01
   1      1.464456230201E-01    2.683183812375E-01   -1.867609847448E-02
   1     -1.295134106348E-01    4.660886106326E-01    3.176497958749E-01
```

**Cellulose I$_\beta$ (100) bilayer slab**

```
1
8.15010000 10.41600000
84
   1     -3.493306357199E-01    2.036847076699E-01    3.146063957894E+00
   1     -3.689396025967E-01   -3.591191357603E-01    3.205365813525E+00
   1      3.147458631753E-01   -4.141971253483E-01    3.150016243345E+00
   1      1.568576569285E-01    7.088983964133E-02    3.168308392435E+00
   1      4.177097327899E-01   -1.808528181809E-01    3.000969180215E+00
   1      4.451670038500E-01    2.248420645710E-01    2.973110428631E+00
   8     -4.626521400246E-01    4.200246225368E-01    2.357166915331E+00
   8     -1.248042860466E-01   -4.945161488696E-01    2.252380870619E+00
   1     -5.469035287661E-03   -4.863871836772E-01    2.080376156544E+00
   6     -3.736489304801E-01   -3.643952871226E-01    2.101003712515E+00
   6     -3.505502210283E-01    2.133843654449E-01    2.042644123203E+00
   8      3.530186421494E-01    2.567573018214E-01    2.137656067915E+00
   6      3.228773146517E-01   -4.203668785260E-01    2.046773997865E+00
   6      1.413466957177E-01    1.003964817547E-01    2.111651621408E+00
   8     -2.616519990038E-01   -5.492702058172E-03    1.858742353931E+00
   8      1.341923573273E-01   -2.446612645860E-01    1.931383817962E+00
   1      6.316692368732E-02    1.865221947926E-01    2.086690269654E+00
   6      4.233298568020E-01   -1.952304629997E-01    1.901157969292E+00
   6      4.433835346660E-01    3.719069235583E-02    1.877515501481E+00
   1     -1.840455159847E-01    3.504301392404E-01    1.891573794032E+00
   1      2.354999681483E-01    4.036422654169E-01    1.864984518787E+00
   1     -2.985720387819E-01   -9.349276708059E-02    1.648971883576E+00
   1      1.187150393779E-01   -1.526704358533E-01    1.667463337187E+00
   6      4.968238475527E-01   -4.633818933977E-01    1.674952637553E+00
   6     -4.902918103474E-01    3.054979696909E-01    1.677493263376E+00
   1     -1.286563044461E-01   -3.054005761966E-01    1.629314484060E+00
   8     -2.003613942344E-01    2.610926092224E-01    1.547539820578E+00
   8      2.028288155467E-01    4.904493474253E-01    1.556512810986E+00
   6     -2.032689064125E-01   -3.930078296473E-01    1.538551270029E+00
   6     -3.857342185006E-01    8.206729138719E-02    1.431082510181E+00
   8     -4.165038639758E-01   -2.417051887637E-01    1.535927965942E+00
   6      2.877024023365E-01   -2.881836084759E-01    1.459610202688E+00
   6      3.092214436336E-01    1.355272742472E-01    1.532522316876E+00
   1     -5.430320723877E-02    1.199038299250E-02    1.411248875651E+00
   8      6.646767335536E-01    1.422553923441E-03    1.340738569348E+00
   8      4.002491402511E-01   -8.037410691597E-02    1.217885237289E+00
   1     -4.941506414268E-01   -4.770563884052E-01    5.816633557719E-01
   1     -4.949360326207E-01    3.193247601145E-01    5.822701475222E-01
   1     -2.187193853641E-01   -4.172161629253E-01    4.707770374771E-01
   1     -3.856612239608E-01    9.148614885265E-02    3.288121924419E-01
   1      2.994742695322E-01    1.458749062458E-01    4.385966100082E-01
   1      2.929671390647E-01   -2.937030054007E-01    3.554521909420E-01
   1     -1.197939582296E-01    3.219300000000E-01   -4.728682754672E-01
   1     -2.938654939344E-01   -1.902400000000E-01   -5.292913996991E-01
   1      1.931008515697E-01    3.869900000000E-01   -6.312779673486E-01
   1      2.174604655873E-01   -5.496000000000E-02   -6.362356477204E-01
   1     -2.162247749528E-04   -2.399900000000E-01   -7.221687741658E-01
   1     -4.514286637897E-03   -4.443300000000E-01   -7.339727750511E-01
   8      9.670360573314E-02    1.580800000000E-01   -1.403259625251E+00
   8      4.310142737118E-01    2.445700000000E-01   -1.504774032865E+00
   8     -9.053197867488E-02   -5.520000000000E-03   -1.530034594759E+00
   8     -3.046991399121E-01    4.984600000000E-01   -1.530270674777E+00
   6     -3.061400323130E-01   -1.586400000000E-01   -1.578430998389E+00
   6     -1.198728710757E-01    3.183300000000E-01   -1.578194918371E+00
   1     -3.818713189211E-01   -7.103000000000E-02   -1.594956599629E+00
   1     -4.488780092833E-01    2.506200000000E-01   -1.680889726074E+00
   6      1.837625820276E-01    3.746900000000E-01   -1.728341809633E+00
   1     -2.082361915822E-01    1.421700000000E-01   -1.730938689828E+00
   6      2.060134224354E-01   -4.921000000000E-02   -1.735896370200E+00
   6     -3.765849367991E-03   -2.249500000000E-01   -1.818288296380E+00
```

```
6    -1.686823370545E-02   -4.573200000000E-01   -1.830564457300E+00
8     2.894883526706E-01   -2.710800000000E-01   -1.856061099213E+00
1    -3.261994436459E-01   -4.143300000000E-01   -1.901152382595E+00
1     3.728176053803E-01    8.567000000000E-02   -1.975753668190E+00
8    -2.428701909361E-01    2.289200000000E-01   -2.020844951572E+00
6     6.348355667718E-02    4.268000000000E-02   -2.046105513467E+00
6     5.038117233971E-02    2.750500000000E-01   -2.058381674388E+00
6    -1.593952607009E-01    4.507900000000E-01   -2.141009680585E+00
1     2.548543533167E-01   -3.578300000000E-01   -2.145967360957E+00
6    -1.371472590559E-01   -1.253100000000E-01   -2.148328161134E+00
1     4.954933322550E-01   -2.493800000000E-01   -2.195780244693E+00
1     4.284866418928E-01    4.289700000000E-01   -2.281713371138E+00
6     1.664910328102E-01   -1.816700000000E-01   -2.298711132413E+00
6     3.527553552847E-01    3.413600000000E-01   -2.298238972378E+00
8     3.513173016466E-01   -1.540000000000E-03   -2.346635376008E+00
8     1.371501404093E-01    4.944800000000E-01   -2.346871456026E+00
8    -3.843961119774E-01   -2.554300000000E-01   -2.372132017920E+00
8    -5.008828276142E-02   -3.419200000000E-01   -2.473410345517E+00
1     5.112960960962E-02    5.567000000000E-02   -3.142697195716E+00
1     4.683438650942E-02    2.610100000000E-01   -3.154737276619E+00
1    -1.708451426156E-01    4.450400000000E-01   -3.240434323047E+00
1    -1.464855285980E-01   -1.130100000000E-01   -3.245392003419E+00
1     3.404808169062E-01    3.097600000000E-01   -3.347378571068E+00
1     1.664092812013E-01   -1.780700000000E-01   -3.403801695300E+00
```

**Cellulose Iβ (100) bilayer slab with aniline**

```
1
8.11110000 10.40480000
94
1    -4.337576600244E-01    3.128544915615E-01    3.828656083886E+00
1    -2.890561091887E-01   -4.851702604296E-01    3.148935199524E+00
6    -3.208678972994E-01    3.084209114802E-01    3.236968497043E+00
1    -3.197365584792E-01    1.012428772443E-01    3.108810947920E+00
6    -2.394794057005E-01    4.212658768382E-01    2.865111483370E+00
6    -2.569811445293E-01    1.901498699123E-01    2.830394418620E+00
6    -9.574346719724E-02    4.156497937593E-01    2.108522891253E+00
6    -1.139172034498E-01    1.828491156228E-01    2.059944008846E+00
1    -3.690927732063E-02   -4.954794954050E-01    1.778737956883E+00
6    -3.245974860960E-02    2.964862671605E-01    1.697275440654E+00
1    -6.685030929957E-02    9.146297006900E-02    1.704186830497E+00
1    -3.568267259535E-01   -2.509927919767E-01    6.900726305261E-01
1    -3.333881288015E-01    3.162771739230E-01    5.898723653743E-01
1     3.235939950348E-01   -3.013447479673E-01    5.054893434950E-01
1     4.179074537684E-01   -7.352317477300E-02    4.495717367808E-01
7     1.110275880988E-01    2.983105506428E-01    9.020026834555E-01
1     4.815219199150E-01    1.252753087131E-01    4.895502566080E-01
1     1.701127273817E-01    3.843292142862E-01    8.254693206708E-01
8    -4.456822378802E-01   -4.674606508312E-01   -2.063796674263E-01
8    -1.030910339003E-01   -3.758606967484E-01   -2.737494484140E-01
6    -3.589093669235E-01   -2.517085853733E-01   -4.173096617774E-01
1     1.404851404003E-02   -3.706935457746E-01   -5.502200642617E-01
6    -3.340354459013E-01    3.260127675773E-01   -5.101310203437E-01
6     3.384725649339E-01   -3.089937913672E-01   -5.929643667187E-01
8    -2.413389102706E-01    1.105515270627E-01   -8.102566922318E-01
8     3.673746561262E-01    3.720919337438E-01   -4.090763545951E-01
6     1.738188715105E-01    2.035155486105E-01    1.219052341931E-01
1    -1.690433401756E-01    4.634998700932E-01   -6.252641152167E-01
6     4.343826390360E-01   -8.389968778706E-02   -6.484797698967E-01
6     4.639210149660E-01    1.472020780681E-01   -5.821508083747E-01
8     1.441604382247E-01   -1.384407586614E-01   -9.813281063795E-01
1     2.539402100790E-01   -4.856825073585E-01   -8.470442611017E-01
1    -2.770186831127E-01    2.350165231092E-02   -1.045120413357E+00
1    -1.160113800320E-01   -1.853499808552E-01   -8.580665318278E-01
8    -1.834618132464E-01    3.753762863366E-01   -9.904042239550E-01
6    -4.850360756019E-01   -3.513373232218E-01   -8.979462258960E-01
6    -4.727104445890E-01    4.180778167625E-01   -8.849257250221E-01
1     1.426143174214E-01   -4.614915625199E-02   -7.265104970715E-01
6    -1.861435852379E-01   -2.749977466392E-01   -9.678378543377E-01
8    -4.020226471457E-01   -1.277195908393E-01   -9.349118178857E-01
```

```
8     2.197314981359E-01  -3.981779425055E-01  -1.117146899879E+00
6    -3.701903740242E-01   1.956371114894E-01  -1.134781398842E+00
6     3.077644025880E-01  -1.778243656599E-01  -1.225411503932E+00
6     3.203906166918E-01   2.435659063938E-01  -7.683369646314E-01
8     1.232655753225E-01   9.175061400790E-02   1.242719652637E-01
8     4.159600034822E-01   3.342650884681E-02  -1.294133198484E+00
1    -4.699827614746E-01  -3.648478521391E-01  -1.986526601980E+00
1    -4.781443564067E-01   4.312686635807E-01  -1.980027021682E+00
1    -1.982486532947E-01  -2.980150305540E-01  -2.041357941892E+00
1    -3.835664851585E-01   2.078775898159E-01  -2.230458009010E+00
1     3.315573974615E-01  -1.847749646462E-01  -2.310841148270E+00
1     2.801272789330E-01   2.394298562229E-01  -1.823671865440E+00
1    -9.163942502443E-02   4.065500000000E-01  -2.931564159044E+00
1    -2.726870812309E-01  -9.559000000000E-02  -2.978960636442E+00
1     2.187911167350E-01   4.727500000000E-01  -3.142777587448E+00
1     2.441962388964E-01   3.125000000000E-02  -3.190404144833E+00
1     2.228293034946E-02  -3.585300000000E-01  -3.200067504302E+00
1     2.646516791563E-02  -1.503000000000E-01  -3.206049583974E+00
8     1.224787342812E-01   2.439900000000E-01  -3.946677063310E+00
8    -2.819429093696E-01  -4.159400000000E-01  -3.955650182817E+00
8    -6.646937774523E-02   8.125000000000E-02  -4.032957058573E+00
6    -9.604158344320E-02   4.034900000000E-01  -4.039399298219E+00
6    -2.851865568864E-01  -7.064000000000E-02  -4.049752897651E+00
8     4.542452248599E-01   3.281500000000E-01  -4.094618495187E+00
1    -3.627913816113E-01   1.640000000000E-02  -4.118086653899E+00
1    -1.834906496003E-01   2.262000000000E-01  -4.186880570122E+00
1    -4.236283270523E-01   3.345100000000E-01  -4.225994167974E+00
6     2.080228289104E-01   4.617800000000E-01  -4.243480247014E+00
6     2.309890645436E-01   3.671000000000E-02  -4.292027124349E+00
6     6.454080477654E-03  -3.708700000000E-01  -4.298239284008E+00
6     1.881243682691E-02  -1.386800000000E-01  -4.307212403515E+00
1    -3.045325578617E-01  -3.276200000000E-01  -4.309283123401E+00
8     3.123905073200E-01  -1.861700000000E-01  -4.421332077249E+00
8    -2.195721860660E-01   3.138300000000E-01  -4.454233515443E+00
1     3.973508791157E-01   1.723800000000E-01  -4.566282469291E+00
6     7.400588442706E-02   3.613200000000E-01  -4.568353189177E+00
6     8.636424077631E-02   1.291300000000E-01  -4.577326308685E+00
6    -1.381707432896E-01  -4.632900000000E-01  -4.583538468343E+00
6    -1.152045076564E-01  -3.822000000000E-02  -4.632085345678E+00
1    -4.835533516938E-01  -1.654900000000E-01  -4.649571424718E+00
1     2.763089708543E-01  -2.738000000000E-01  -4.688685022570E+00
1     4.556097028652E-01  -4.836000000000E-01  -4.757478938793E+00
8    -3.614241252380E-01  -1.718500000000E-01  -4.781177177492E+00
6     3.780048781403E-01   4.293600000000E-01  -4.825812695042E+00
6     1.888599046972E-01  -9.651000000000E-02  -4.836166294473E+00
8     1.592876989992E-01  -4.187500000000E-01  -4.842608534119E+00
8     3.747640089915E-01   8.406000000000E-02  -4.920145489862E+00
8    -2.965763465926E-02  -2.560100000000E-01  -4.929118609370E+00
1     6.635315333834E-02   3.497000000000E-01  -5.669516008719E+00
1     7.053539090451E-02   1.414700000000E-01  -5.675498088390E+00
1    -1.513779176424E-01  -4.687500000000E-01  -5.685161447859E+00
1    -1.259701171131E-01  -2.725000000000E-02  -5.733018085232E+00
1     3.655054024849E-01   4.044100000000E-01  -5.896604956250E+00
1     1.844577462784E-01  -9.345000000000E-02  -5.944001433648E+00
```

**Cellulose Iβ (100) bilayer slab with 4,4'diaminoazobenzene**

```
1
8.11110000 10.40480000
108
1    -1.588936475183E-01  -4.205846489003E-01   2.638251372953E+00
1    -2.248935254346E-01   4.454416103551E-01   1.812299783717E+00
7    -1.312054064328E-01   4.893422300400E-01   2.322413185297E+00
1     1.241517129822E-01  -3.585206229473E-01   2.860308174240E+00
6     2.956708249431E-02   4.648718413042E-01   2.000834474306E+00
1    -2.203289933105E-02   2.804211621075E-01   1.076639521028E+00
6     1.562131507424E-01  -4.471805124980E-01   2.335383855350E+00
6     7.412071168099E-02   3.493255350137E-01   1.336946460931E+00
6     3.178011487677E-01  -4.711080819047E-01   1.971563999770E+00
6     2.350615855732E-01   3.267121552626E-01   9.683041052975E-01
```

```
 1      4.150960910876E-01  -4.018843603828E-01   2.196682694907E+00
 6      3.605888422463E-01   4.167646735026E-01   1.262392647970E+00
 1      2.687358451100E-01   2.403583672127E-01   4.150129115097E-01
 7     -4.764515318442E-01   4.065103364018E-01   8.289297776351E-01
 7     -4.431008198600E-01   3.054192401212E-01   1.489052732116E-01
 1     -2.609607644514E-01  -4.873552993189E-01  -5.862089053435E-01
 6     -2.872866407988E-01   3.067450592328E-01  -4.534054788491E-01
 1     -2.859747154607E-01   1.006092329501E-01  -5.599457259111E-01
 6     -2.037372850879E-01   4.205777673517E-01  -8.157708632884E-01
 6     -2.187286956405E-01   1.878512342239E-01  -8.237990892181E-01
 6     -5.575969109806E-02   4.143819108654E-01  -1.500657229263E+00
 6     -7.020581050948E-02   1.810048520990E-01  -1.512779366694E+00
 1      3.947385896447E-03  -4.969143436816E-01  -1.822875670781E+00
 6      1.191934903817E-02   2.949583591981E-01  -1.864003907535E+00
 1     -1.957115802725E-02   8.947528586615E-02  -1.830493018375E+00
 1     -3.125928016461E-01  -2.523866362471E-01  -2.858624916043E+00
 1     -2.897493495003E-01   3.143808465070E-01  -2.987463163715E+00
 1      3.683398647323E-01  -3.034323214283E-01  -3.038173514302E+00
 1      4.615358934328E-01  -7.443834853317E-02  -3.088896177998E+00
 7      1.586348191488E-01   2.963257277375E-01  -2.610649901462E+00
 1     -4.739504897855E-01   1.252539207599E-01  -3.061793872808E+00
 1      2.223753929578E-01   3.806491755613E-01  -2.632414065152E+00
 8     -4.003978781840E-01  -4.686178476979E-01  -3.766386218933E+00
 8     -5.937052029312E-02  -3.772411916592E-01  -3.850903745058E+00
 6     -3.149078610806E-01  -2.525498471682E-01  -3.965898081081E+00
 1      5.805464721895E-02  -3.715386566244E-01  -4.118919369523E+00
 6     -2.901969213853E-01   3.246181634239E-01  -4.084666716980E+00
 6      3.827786332508E-01  -3.104983590480E-01  -4.137282847971E+00
 8     -1.979774555996E-01   1.093159878625E-01  -4.386225268137E+00
 8      4.120965810389E-01   3.711984293522E-01  -3.952668051362E+00
 6      2.174898350284E-01   2.032009021941E-01  -3.438451926673E+00
 1     -1.253002927556E-01   4.621300747803E-01  -4.200443070248E+00
 6      4.777569924070E-01  -8.496982148153E-02  -4.186998263920E+00
 6     -4.923823445322E-01   1.462679986437E-01  -4.134613810463E+00
 8      1.878191160971E-01  -1.404347390418E-01  -4.505602431027E+00
 1      2.973548852402E-01  -4.870034290441E-01  -4.379446311323E+00
 1     -2.342877788510E-01   2.204287179777E-02  -4.605439022064E+00
 1     -7.197844912386E-02  -1.855848839393E-01  -4.391761544150E+00
 8     -1.391776437723E-01   3.736418975367E-01  -4.557981653443E+00
 6     -4.406890858725E-01  -3.522562151172E-01  -4.449792664582E+00
 6     -4.288066163850E-01   4.171840087948E-01  -4.445404732359E+00
 1      1.852139806717E-01  -4.707124711913E-02  -4.301633835432E+00
 6     -1.422239314020E-01  -2.749129910446E-01  -4.521239326277E+00
 8     -3.585899443082E-01  -1.283937241791E-01  -4.476291252070E+00
 8      2.641794499209E-01  -3.996520475126E-01  -4.663440986416E+00
 6     -3.271440551741E-01   1.942321271858E-01  -4.704918095684E+00
 6      3.513524285922E-01  -1.791533006547E-01  -4.762534600123E+00
 6      3.645287190415E-01   2.431035381256E-01  -4.318390458298E+00
 8      1.616385182722E-01   9.326646044512E-02  -3.488361007437E+00
 8      4.585228976325E-01   3.217989410671E-02  -4.835558682324E+00
 1     -4.259657891315E-01  -3.651762266134E-01  -5.539386338156E+00
 1     -4.361722867665E-01   4.306948735965E-01  -5.539211194576E+00
 1     -1.545041301649E-01  -2.956730998267E-01  -5.599593245983E+00
 1     -3.419955478251E-01   2.059444199890E-01  -5.799935755922E+00
 1      3.745903301808E-01  -1.855164824596E-01  -5.849497468953E+00
 1      3.242582658852E-01   2.394575510576E-01  -5.374033765155E+00
 1     -4.877059705398E-02   4.065500000000E-01  -6.481583324132E+00
 1     -2.298182532605E-01  -9.559000000000E-02  -6.528979801530E+00
 1      2.616599447054E-01   4.727500000000E-01  -6.692796752536E+00
 1      2.870650668669E-01   3.125000000000E-02  -6.740423309921E+00
 1      6.515175831991E-02  -3.585300000000E-01  -6.750086669390E+00
 1      6.933399588608E-02  -1.503000000000E-01  -6.756068749062E+00
 8      1.653475622516E-01   2.439900000000E-01  -7.496696228398E+00
 8     -2.390740813991E-01  -4.159400000000E-01  -7.505669347906E+00
 8     -2.360054977478E-02   8.125000000000E-02  -7.582976223661E+00
 6     -5.317275547275E-02   4.034900000000E-01  -7.589418463307E+00
 6     -2.423177289159E-01  -7.064000000000E-02  -7.599772062739E+00
 8      4.971140528304E-01   3.281500000000E-01  -7.644637660276E+00
 1     -3.199225536408E-01   1.640000000000E-02  -7.668105818987E+00
 1     -1.406218216298E-01   2.262000000000E-01  -7.736899735210E+00
 1     -3.807594990818E-01   3.345100000000E-01  -7.776013333063E+00
```

```
6       2.508916568808E-01    4.617800000000E-01   -7.793499412102E+00
6       2.738578925140E-01    3.671000000000E-02   -7.842046289437E+00
6       4.932290844810E-02   -3.708700000000E-01   -7.848258449096E+00
6       6.168126479736E-02   -1.386800000000E-01   -7.857231568603E+00
1      -2.616637298913E-01   -3.276200000000E-01   -7.859302288490E+00
8       3.552593352905E-01   -1.861700000000E-01   -7.971351242338E+00
8      -1.767033580956E-01    3.138300000000E-01   -8.004252680531E+00
1       4.402197070861E-01    1.723800000000E-01   -8.116301634379E+00
6       1.168747123975E-01    3.613200000000E-01   -8.118372354265E+00
6       1.292330687468E-01    1.291300000000E-01   -8.127345473773E+00
6      -9.530191531914E-02   -4.632900000000E-01   -8.133557633432E+00
6      -7.233567968598E-02   -3.822000000000E-02   -8.182104510766E+00
1      -4.406845237233E-01   -1.654900000000E-01   -8.199590589806E+00
1       3.191777988247E-01   -2.738000000000E-01   -8.238704187659E+00
1       4.984785308357E-01   -4.836000000000E-01   -8.307498103882E+00
8      -3.185552972676E-01   -1.718500000000E-01   -8.331196342580E+00
6       4.208737061108E-01    4.293600000000E-01   -8.375831860130E+00
6       2.317287326676E-01   -9.651000000000E-02   -8.386185459561E+00
8       2.021565269696E-01   -4.187500000000E-01   -8.392627699208E+00
8       4.176328369619E-01    8.406000000000E-02   -8.470164654951E+00
8       1.321119331119E-02   -2.560100000000E-01   -8.479137774458E+00
1       1.092219813088E-01    3.497000000000E-01   -9.219535173807E+00
1       1.134042188750E-01    1.414700000000E-01   -9.225517253478E+00
1      -1.085090896720E-01   -4.687500000000E-01   -9.235180612948E+00
1      -8.310118914263E-02   -2.725000000000E-02   -9.283037250320E+00
1       4.083742304553E-01    4.044100000000E-01   -9.446624121339E+00
1       2.273265742489E-01   -9.345000000000E-02   -9.494020598736E+00
```